\documentclass[12pt]{article}
\usepackage{amsmath,amssymb}
\usepackage{graphicx}
\usepackage[dvips]{color}
\usepackage[dvips]{color}
\usepackage{graphicx}
\usepackage{epsf}
\usepackage[dvips]{color}
\usepackage{graphicx}
\usepackage{epsf}
\usepackage{amsmath}
\usepackage{amssymb}\usepackage{amsfonts}
\input amssym.def
\input amssym.tex
\def \nn{\nonumber}

\def \const {{\rm constant}}
\def \cE {{\cal E}}

\makeatletter
\@addtoreset{equation}{section}
\makeatother

\newcommand{\ba}{\begin{eqnarray}}
\newcommand{\ea}{\end{eqnarray}}
\def\half{\frac{1}{2}}
\def\ben{\begin{equation}}
\def\een{\end{equation}}
\def\bea{\begin{eqnarray}}
\def\eea{\end{eqnarray}}
\def\be{\begin{equation}}
\def\ee{\end{equation}}

\def\nn{\nonumber}
\def\p{\partial}

\def\ben{\begin{equation}}
\def\een{\end{equation}}
\def\bea{\begin{eqnarray}}
\def\eea{\end{eqnarray}}
\def \nn {\nonumber}

\def \bx {{\bf x}}
\def \p {\partial}

\def\half {\frac{1}{2}}

\newcount\hour \newcount\minute
\hour=\time  \divide \hour by 60
\minute=\time
\loop \ifnum \minute > 59 \advance \minute by -60 \repeat
\def\nowtwelve{\ifnum \hour<13 \number\hour:% 		% supresses leading 0's
                      \ifnum \minute<10 0\fi%		% so add it it
                      \number\minute
                      \ifnum \hour<12 \ A.M.\else \ P.M.\fi
	 \else \advance \hour by -12 \number\hour:% 	% supresses leading 0's
                      \ifnum \minute<10 0\fi%		% add it in
                      \number\minute \ P.M.\fi}
\def\nowtwentyfour{\ifnum \hour<10 0\fi% 		% need a leading 0 
		\number\hour:% 				% supresses leading 0's
         	\ifnum \minute<10 0\fi% 		% add it in
         	\number\minute}

\newcommand{\hoch}[1]{$\, ^{#1}$}

\newcommand{\auth}{\Large\bf{ 
G.W. Gibbons\hoch{1,2,3,4}}}

\begin{document}

%\begin{flushright}
%\hfill {
%UPR-1273-T\ \ \ DAMTP-2015-125\ \ \
%MIFPA-14-29 \ \ \
%}\\
%\end{flushright}

\begin{center}

{\LARGE{\bf The Jacobi-metric for timelike geodesics
in static spacetimes}}

\vspace{10pt}
\auth

\large

\vspace{3pt}{\hoch{1}\it DAMTP, Centre for Mathematical Sciences,\\
 Cambridge University, Wilberforce Road, Cambridge CB3 OWA, UK}

\vspace{7pt}{\hoch{2}\it Department of Physics and Astronomy,\\
University of Pennsylvania, Philadelphia, PA 19104, USA}

%\vspace{3pt}{\hoch{2}\it DAMTP, Centre for Mathematical Sciences,\\
% Cambridge University, Wilberforce Road, Cambridge CB3 OWA, UK}

\vspace{3pt}{\hoch{3}\it  
Laboratoire de Math\'ematiques et Physique Th\'eorique  CNRS-UMR
7350 \\
F\'ed\'eration Denis Poisson, Universit\'e Fran\c cois-Rabelais Tours,\\
Parc de Grandmont, 37200 Tours, France}                                         

\vspace{3pt}{\hoch{4}\it 
LE STUDIUM, Loire Valley Institute for Advanced Studies,\\
Tours and Orleans, France}

%\vspace{3pt}{\hoch{5}\it George P. \& Cynthia W. Mitchell
%Institute for\\ 
%Fundamental Physics and Astronomy,\\ Texas A\&M University,
%College Station, TX 77843-4242, USA}

%\vspace{3pt}{\hoch{6}\it %Department of Physics and Astronomy,\\
% Center for Applied Mathematics and Theoretical Physics,
% University of Maribor, SI2000 Maribor, Slovenia}
\vskip .5 cm
%\centerline{$\frak{DRAFT}$}
%\centerline{\today\,\,\now}
\vskip .5cm 
\begin{abstract}
{\textnormal{It is shown that the free motion of massive particles moving
in static spacetimes are given by the  geodesics of an energy-dependent
Riemannian metric on the spatial sections analogous to Jacobi's metric
in classical dynamics. In the massless limit Jacobi's metric
coincides with  the energy independent 
Fermat or optical metric.  For  stationary metrics,  it is known that
the motion of massless particles  is given by the geodesics of 
an energy independent
Finslerian metric of Randers type. The motion of massive particles
is governed by neither a Riemannian nor a Finslerian metric. 
The properies of the Jacobi metric for massive particles 
moving outside the horizon of a Schwarschild black hole are
described. By constrast  with the massless case, the
Gaussian  curvature of the equatorial sections is not always  negative. }}    

\end{abstract}
\end{center}

\pagebreak    

\tableofcontents

\pagebreak
\section{Introduction}

An elegant device for implementing the 
Principle of Least Action of Maupertuis was introduced by Jacobi. 
One varies the action of a mechanical system 
\ben \int_\gamma  p_i dx^i = \int_\gamma  p_i \dot x^i dt \een  
along an unparameterized   path $\gamma$ in an $n$ -dimensional   configuration space
$Q$ with coordinates $x^i\,,i=1,2,\dots,n$ and canonical momenta $p_i$ 
subject to the constraint that along the curve $\gamma$ the energy $E$
is conserved. An equivalent formulation is to lift the curve $\gamma$
to the cotangent space $T^\star Q$ and restrict variations to  a level
set of the Hamiltonian $H(x,p)=E$.

In the simplest case  the 
kinetic energy $T= \half m_{ij}(x)  \dot x ^i \dot x^j$ ,
where the space dependent  mass matrix   $ m_{ij}(x) dx^ dx^j$ 
endows the  configuration space $Q$ with a  Riemannian metric
and the Lagrangian giving the equations  of motion  is   
\ben
L= \half m_{ij}(x)  \dot x ^i \dot x^j - V(x)  \,.  
\een
Jacobi showed that the unparamaterised curves
extremizing the constrained  action are geodesics
of the rescaled  Jacobi metric
\ben
j_{ij} (E,x) dx^idx^j = 2(E-V) m_{ij} dx^idx^j \,.  
\een  
One recovers the parameterization of the motion as   a function
of physical time $t$ by noting that length $s$ with respect to the 
Jacobi metric
is related to the  original  time parameter 
$t$ by
\ben
dt =\frac{ds}{2(E-V)} \,. \label{time}
\een    

This procedure  opened up  the way open to investigations of the motion
of the original mechanical system using the  methods
developed by differential geometers to investigate geodesic motion. 
Of particular interest is the influence of the
curvature of the Jacobi metric \cite{Ong,Pettini}.   
An important application to gravity was
the work by Ong \cite{Ong} who studied the curvature of the
the Jacobi metric for the  Newtonian $N$-body problem (see also \cite{Savvidy}).
One has  $Q=\mathbb{R}^{3n}$ with the flat metric 
\ben
m_{ij}dx^i dx^j = \sum_a m_a d \bx_a^2 \,,
\een
and the potential energy is 
\ben
V= - \sum _{1 \le a < b \le  N}   \frac{Gm_am_b}{|\bx_a-\bx_b|} \,. 
\een
Since  in this case, the  Jacobi metric is conformally flat
the evalution of the curvature is straight forward.
If $ N=2$, the problem reduces to the Kepler's problem of
the relative motion and the relevant Jacobi metric is
up to an unimportant over all constant factor \footnote{Note that $E$
in this Newtonian case does not contain a contribution from
the rest mass of the particle.}. 
\ben
(\frac{E}{m} + \frac{M}{r} ) 
\left(dr ^2 +r^2 (d\theta ^2 + \sin ^2 \theta d \phi)^2\right. 
\een
By symmetry, one may restrict attention to the equatorial
plane $\theta = \frac{\pi}{2}$ which is a totally geodesic 
submanifold. One then has a $2$ -dimensional  axially symmetric   
metric  
\ben
(\frac{E}{m} + \frac{M}{r} ) 
(dr ^2 +r^2  d \phi ^2 )\,. \label{axi2} 
\een
Ong \cite{Ong} showed that the sign of the   Gaussian  
curvature of the metric has the opposite  sign to that of the energy $E$.
If $E > 0$, which of course corresponds  
to unbound hyperbolic or parabolic  orbits, 
he showed that the  Jacobi metric (\ref{axi2}) 
is well defined and complete
for  $ 0 \le r < \infty$ and  if $E<0$, which corresponds to bound 
elliptical orbits,  
it is well defined  for $0< r< \frac{2Mm}{E}$. 

Ong also gave an isometric embedding of the Jacobi manifold
into three dimensional Euclidean space $\mathbb{E}^3$ 
with Cartesian coordinates $x,y,z$ 
as a  surface of revolution $z= f(\sqrt{x^2 + y^2}) $.   
If $E=0$, then $f^{\prime \prime} =0$ and the surface is the  cone
$z=\sqrt{3} \sqrt{x^2+y^2} $ which has  
deficit angle $\pi$, or equivalently, semi-angle $30^\circ$. 
In the other two  cases the surface approaches   the  cone near the origin. 
If $E>0$, then  $ f^{\prime \prime} <0$ ; the surface has  negative 
Gauss curvature and remains outside the cone . If 
$E<0$, then $ f^{\prime \prime} >0$   the  surface remains inside the cone and 
asymptotes the cylinder $\sqrt{x^2 + y^2}  = \frac{2Mm}{-E}$.
Ong also studied the three body problem using these techniques.

It is obviously of interest, if only to extend one's intuition
by means of an easily visualised model,
to see whether these ideas can be applied to General Relativity.
At a formal level, one takes the configuration space $\{Q,m_{ij}\}$ 
to be Wheeler's superspace equipped with its DeWitt metric
(cf. \cite{Baierlein:1962zz}).
In the vacuum case, the potential $V$ is
\ben
\int_\Sigma  R  \sqrt{g} \,d^3\, x  
\een 
and the Hamiltonian constraint implies that the energy vanishes $E=0$.
One then obtains a picture of spacetime as a sheaf of 
geodesics in superspace. Having obtained the geodesic
between two points in superspace one obtains the time
duration between them using (\ref{time}), thus solving the much discussed
``problem of time''. 

Less ambitiously one may confine attention to
a mini-superspace truncation,      and this has been 
done in attempts to investigate inflation \cite{Gurzadyan} 
and  the  chaotic behaviour
of Bianchi IX Mixmaster models (see e.g.\cite{Biesiada:1994nz}).

The focus of the present paper is different. It is the motion
a  test particle
of rest mass $m$ following a timelike geodesic in a stationary
background. A limiting case would be a zero rest-mass particle.
This latter case is well known in the static case
to reduce to geodesic motion with respect
to the optical or  Fermat metric $f_{ij}(x)= \frac{1}{-g_{tt}} g_{ij}$.
This was studied in 
\cite{Gibbons:1993cy,Gibbons:2008rj}     
and a number of limitations
on possible motions using the Gauss-Bonnet theorem.
In particular in \cite{Gibbons:2008rj} the Gaussian curvature of the
Schwarzschild  optical metric restricted to the equstorial plane
was shown  to be 
everywhere negative and to approach a constant value near
the horizon $r=2M$. In \cite{Gibbons:2008hb} 
this behaviour was found to be universal
for the near horizons of non-extreme static black holes.
One purpose of the present paper is to extend this work to
the case of massive particles. The extension of 
Fermat's principle to cover stationary spacetimes 
entails replacing the Riemmanian metric by a
 Finsler metric of Randers
type (see e.g.  \cite{Gibbons:2008zi,Werner:2012rc,Bloomer:2011rd}.

\section{The  Jacobi metric for static spacetimes}

If
\ben
ds =-V^2 dt ^2 + g_{ij}dx^idx^j,
\een
the action for a massive particle is 
\ben
S= -m \int L dt = - m \int dt \sqrt{V^2 - g_{ij}\dot x^i \dot x^j }  
\een
where $\dot x^i= \frac{d x^i}{dt} $
The canonical momentum is 
\ben
p_i= \frac{m \dot x^i}{ \sqrt{V^2 - g_{ij}\dot x^i \dot x^j }} \,.                              
\een
whence the Hamiltonian is
\bea
H&=& \frac{mV^2 }{ \sqrt{ V^2 - g_{ij}\dot x^i \dot x^j }}   \\ 
&=& \sqrt {    m^2  V^2 + V^2 g^{ij} p_i p_j  } \,.
\eea
Setting 
\ben
p_i=\p_i S \,, 
\een
 the Hamilton-Jacobi equation becomes
\ben
\sqrt{m^2 V^2 + V^2 g^{ij}\p_iS \p_j S j  } =E 
\een
or \ben
f^{ij} \p_iS \p_jS  = E^2 - m^2 V^2 \,, 
\een
where $f^{ij}f_{jk}= \delta ^i_k$ and
\ben
f_{ij} = V^{-2} g_{ij} 
\een 
is the optical or Fermat metric.
Thus
\ben
 \frac{1}{E^2 - m^2 V^2  }   f^{ij} \p_iS \p_jS  = 1 \,, 
\een
which is the Hamilton-Jacobi equation for  geodesics of the 
Jacobi-metric $j_{ij}$ given by
\ben
j_{ij}dx ^i dx ^j = \bigl( E^2 -m^2 V^2) V^{-2} g_{ij} dx ^i dx ^j.
\een 
Note that the massles case, $m=0$,  the Jacobi metric coincides with the
Fermat metric up to a factor of $E^2$ and as a  consequence
the geodesics, considered as unparameterized curves, do not depend
upon the energy $E$. However in the massive case, $m\ne 0$  the geodesics
\emph{do} depend upon $E$.   

In general, if the spacetime is asymptotically flat 
and the sources obey the energy conditions, then   $0\le V\le 1$. 
Therefore,  if $E^2\ge m^2$ , the Jacobi metric 
is positive definite and complete, even if horizons are present.
If however $E^2 <m^2$ there are bound orbits  and the Jacobi 
metric changes signature
at large distances. Generically there will be a level set
of $V$ on which   $E^2 -m^2 V^2$ vanishes and hence on which  the Jacobi metric vanishes.
From the point of view of the Jacobi metric this level set
is a point-like  conical singularity. Every geodesic
 must have a turning point   on or inside this level set.

\section{The Schwarzschild Case} 

In the case of the  Schwarzschild solution the Jacobi metric is 
\ben
ds ^2 = \Bigl( E^2-m^2 + \frac{2M m^2 }{r}  \Bigr )
 \Bigl( \frac{dr^2 }{(1- \frac{2M}{r} )^2 } + 
\frac{r^2 }{(1-\frac{2M}{r})  }  
 \bigl(d \theta ^2 + \sin ^2 \theta d \phi ^2 \bigr )  \Bigr ) \,.
\label{SJmetric}
\een
The first large  bracket in (\ref{SJmetric} ) is the conformal factor
and the second large bracket the optical metric.
The later is defined for $2M < r < \infty$ 
and the horizon at $r=2M$ is infinitely far way with respect to the radial
optical radial distance or tortoise coordinate  
\ben
r^\star = \int^r _{2M}  \frac{dx}{1-\frac{2M}{x} } = 
r -2M +  \ln (\frac{r}{2M }-1)\,,\quad \Leftrightarrow \quad 
r-2M=2M W(e^\frac{r^\star}{2M}) \,, 
\een 
where $W(x)$ is Lambert's function defined by $\ln x =W(x) + \ln W(x)$.

By spherical symmetry, in order to study geodesics, it is sufficient to consider
the equatorial plane $\theta=\frac{\pi}{2}$ on which the restriction of
the Jacobi metric is  
 \ben
ds ^2 = \Bigl( E^2-m^2 + \frac{2M m^2 }{r}  \Bigr )
 \Bigl( \frac{dr^2 }{(1- \frac{2M}{r} )^2 } + 
\frac{r^2 }{(1-\frac{2M}{r})}d \phi ^2   \Bigr ) \,. \label{Sjacobi}
\een
By axi-symmetry  we have a conserved quantity often called Clairaut's constant
which corresponds     physically  to angular momentum. That is   
\ben
l= \Bigl( E^2-m^2 + \frac{2M m^2 }{r}  \Bigr ) \frac{r^2} {(1-\frac{2M}{r})}   \frac{d \phi}{ds} =  \const \,.
\een
 Now 
\ben
 \Bigl( E^2-m^2 + \frac{2M m^2 }{r}  \Bigr )
 \Bigl( \bigl ( \frac{dr }{d s } )^2  \frac{1}{(1-2 \frac{M}{r} )^2 } + 
\frac{r^2 }{(1-\frac{2M}{r})} \bigl( \frac{d \phi}{ds} \bigr )^2    
\Bigr ) =1 \,.
\een
whence 
\ben
\Bigl( E^2-m^2 + \frac{2M m^2 }{r}  \Bigr ) ^2    
\frac{1}{(1-2 \frac{M}{r} )^2 }  
\bigl(  \frac{dr }{d s} \bigl )^2 = E^2 - \big(1-\frac{2M}{r} ) 
\bigl(m^2 + \frac{l^2}{r^2} \bigr )  \,, 
\een
which agrees  with the standard result that  
\ben
m^2 (\frac{dr }{d \tau })^2 = E^2 -  \bigl( 1-2 \frac{M}{r}   \bigr )   
 \bigl( m^2 + \frac{l^2}{r^2} \bigr) \,, \label{radial}
\een
where $\tau$ is proper time along the particle's worldline
and  
\ben
l= m  r^2 \frac{d \phi}{d \tau}   
\een
as long as 
\ben
  d \tau = m \frac{ 1-2 \frac{M}{r}} 
{ E^2-m^2 + \frac{2M m^2 }{r} } \, ds  \,.
\een
Thus $l$ is indeed angular momentum.
In standard treatments  $u=\frac{1}{r} $ satisfies  Binet's
equation
\ben
\frac{d^2u}{d \phi ^2 } + u = \frac{F(u)}{h^2 u^2}\, , 
\een
where, for  a massive particle orbiting a  Schwarzschild  black hole, we have
\ben
\frac{F(u)}{h^2 u^2} = 3M u^2 + \frac{M}{h^2}\,, \label{tardyon}
\een
and where $h=\frac{l}{m} $ is the conserved angular momentum per unit mass.
If this were a classical central orbit problem we would
say that   we have a sum of an inverse fourth and inverse square law attraction.
There is a first integral
\ben
\bigl (\frac{du}{d \phi } \bigr) ^2 =- u^2 +2Mu^3 + \frac{2Mu}{h^2} +C 
= 2M(u-\alpha)(u-\beta)(u-\gamma)   
\label{first}\een
where $C$ is  a constant related to the energy per unit mass
$\cE=\frac{E}{m}$   and  the
angular momentum per unit mass $h$  by 
\ben
C= \frac{\cE ^2-1}{h^2} 
\een
with
\ben
\alpha + \beta + \gamma = \frac{1}{2M}\,,\qquad \beta \gamma + \gamma \alpha
+ \alpha  \beta = \frac{1}{h^2} \,,\qquad \alpha \beta \gamma =-\frac{C}{2M} \,. \een
\subsection{Some  Explicit Solutions} 

The behaviour of the orbits depend on the two  dimensionless
quantities which are the specific energy  $\cE >0$ and  $\frac{M}{h}$. 
In general, the solutions are
given by elliptic integrals. However there  is a one parameter  family
of explicit solutions of the form
\bea
u&=&A +  \frac{B}{\cosh^2 (\omega \phi)}\\ 
A &=& \frac{1}{6M} \bigl( 1 \pm \sqrt{1- \frac{12 M^2}{h^2}  }\bigr ) \\ 
B &=& \mp  \frac{1}{2M} \sqrt{1- \frac{12 M^2}{h^2}}\\ 
\omega ^2& =& \pm \frac{1}{4} \sqrt{1- \frac{12 M^2}{h^2}}\\
C&=& A^2 ( 4MA-1) \,.  
\eea
These solutions arise because two of the 
the roots of the cubic eqns in (\ref{first}) 
coincide.

If $\omega$ is real and $A+B\ne 0$   
the solutions are symmetric about $\phi=0\,, r= \frac{1}{A+B}$    
and end  spiralling  around  a circular geodesic  at $r=\frac{1}{A}$. 
If $h^2= 12M^2 $ then $B=0$ and we have the inner most stable circular
orbit at $r=6M$ for which the specific energy $\cE = \sqrt{\frac{8}{9}}$. 
If $h^2= 16 M^2 $  the energy per unit mass $\cE=1$.
Since $A=-B= \frac{1}{4M}$,   these orbits are in free fall
from infinity, starting from rest and spiral around a circular  orbit
at $r=4M$. All orbits starting from rest  at  infinity (i.e. having
$\cE=1$) with     $|h| < 4M$  fall through the horizon at $r=2M$
while all such orbits with $|h| >4M$  are scattererd back to infinty.
These  latter orbits are relevant for the theory of the 
BSW effect \cite{Banados:2009pr}.

\subsection{Bound States and Jacobi functions}

For a bound orbit we have  three real positive roots 
taken to satisfy  
\ben\alpha> \beta \ge  u \ge \gamma >0 \, . \label{range}
\een
The first integral (\ref{first}) leads to
\ben
\half \varpi  d \phi = 
\frac{\sqrt{\alpha-\gamma}\,du }{\sqrt{4(\alpha-u)(\beta-u)
(u-\gamma) }}   \label{integral}
\een
with
\ben
\varpi = \sqrt{2M(\alpha-\gamma)}= 
\sqrt{\frac{\alpha-\gamma}{\alpha + \beta +\gamma}}
\,.
\een
Thus \cite{Greenhill} 
\bea
u &=& \gamma + (\beta-\gamma) {\rm  sn}^2(  \frac{\varpi \phi}{2}) \nn \,,\\ 
 &=& \beta  - (\beta-\gamma) {\rm  cn}^2(  \frac{\varpi \phi}{2}) \nn \,,\\
 &=& \alpha  - (\alpha -\gamma) {\rm  dn}^2(  \frac{\varpi \phi}{2}\,,) 
\label{ellipticintegrals}
\eea
where the modulus $k$  of the elliptic functions is given by
\ben
k= \sqrt{\frac{\beta-\gamma}{\alpha-\gamma}} \,, 
\een
and the quarter period $K$ by
\ben
K= \int_0 ^{\frac{\pi}{2}} \frac{d \theta}{\sqrt{1-k^2 \sin^2 \theta}} \,.  
\een

Using  (\ref{ellipticintegrals}) and properties
of  elliptic functions,   (\ref{integral})  may 
expressed as 
\ben
\frac{1}{r}= \frac{1}{r_p} {\rm cn}  ^2 (\frac{\varpi \phi}{2}) + 
\frac{1}{r_a} {\rm sn}  ^2 (\frac{\varpi \phi}{2})
= \frac{1}{L} \Bigl (1+ e 
[{\rm cn}  ^2 (\frac{\varpi \phi}{2})  - {\rm sn}  ^2 (\frac{\varpi \phi}{2}]    ) \Bigr ) \,,
\een 
where the constants $r_p \le r_a$ are the radii at perihelion
and aphelion respectively, since from (\ref{range}) 
\ben
r_p = \frac{1}{\gamma} \le r \le \frac{1}{\beta} =r_a  
\een
and 
\ben
L=\frac{2r_pr_a}{r_p+r_a}\,,\qquad e=\frac{r_a-r_p}{r_a+r_a} \,.
\een

\bea 
{\rm Perihelion \, is \, at }:  \phi&=& \frac{4K}{\varpi} n\,,\qquad n \in {\Bbb Z} 
\\[6pt]
{\rm Aphelion \, is \, at } :  \phi&=& \frac{4K}{\varpi} (n+\half ) 
\,, \qquad n \in {\Bbb Z}
\eea

These expressions generalise the Newton-Kepler case for which the orbits 
are ellipses with foci at the origin and  are given by
\ben
\frac{1}{r}= \frac{1}{r_p} \cos^2 (\frac{\phi}{2}) + 
\frac{1}{r_a} \sin^2 (\frac{\phi}{2}) = \frac{1}{L} (1+ e\cos\phi)\,,
\een
where  $L$ is the semi-latus rectum
and $e$ is the eccentricity Note that in both cases $L$ is
the harmonic mean of the perihelion and aphelion radii.
 
\subsection{Relation to Weierstrass Functions and Photon Orbits}
The term linear in $u$ may be eliminated by setting
\ben
u=v+c\,,
\een
where
\ben
c^2 - \frac{c}{3M} + \frac{1}{3h^2} \, ,\, \Longleftrightarrow
\, c= \frac{1}{6M} \bigl(1\pm \sqrt{ 1- \frac{4M^2}{h^2} }  \bigr)\,.
 \, \Longleftrightarrow
\, 1- 6Mc = \mp \sqrt{ 1- \frac{4M^2}{h^2} } \,,
\een
We then find that if $\tilde \phi= \phi \sqrt{1-6Mc}$, 
$\tilde M = \frac{M}{1-6Mc}$, and $\tilde C = C- c^2(1-2Mc)$,   then
\ben
\bigl (\frac{dv}{d \tilde \phi } \bigr) ^2 =- v^2 +2\tilde Mv^3 + \tilde C,
\een 
which is the equation governing the orbit of a photon
in a Schwarzschild solution of mass $\tilde M$. 
Thus the solution is \cite{Gibbons:2011rh} 
\ben
\tilde M v= \frac{1}{6} + 2 \frak{p} (\tilde \phi), 
\een
 where $\mathfrak{p}$ is the  Weierstrass's Elliptic function with parameter 
$g_3= \frac{1}{216} - \bigl(\frac{\tilde M}{2}\bigr)^2 \tilde C $.

It follows that the massive particle orbits are given by
\ben
u=\frac{1}{6  M} + \frac{1-6Mc} {3 M} \frak{p} 
(\phi\sqrt{1-6Mc}) \,.
\een

A particular example is provided by the 
cardioidal photon orbit \cite{Gibbons:2011rh} 
with $\tilde C=0$, 
\ben
v= \frac{1}{2\tilde M \cos^2 (\frac{\tilde\phi}{2})}, 
\een
which gives  
\ben
 u=c + v= c + \frac{1-6Mc}{2M\cos ^2 (\frac{\sqrt{1-6Mc}\phi} {2})}. 
\een
If we take the case for which $1-6Mc >0$ this gives
\ben
6M u= 1- \sqrt{1-\frac{4M^2}{h^2} } + 
3 \frac{\sqrt{1-\frac{4M^2}{h^2} }}
{ \cos ^2 (\frac{ ( 1-\frac{4M^2}{h^2}) ^{\frac{1}{4}}  \phi} {2}  )} \,.
\een 
These orbits run  from the past singularity at $r=0$  out to
a maximum radius $r_{\rm max}$ given by
\ben
r_{\rm max} = \frac{6M}{1 + 2 \sqrt{1-\frac{4M^2}{h^2} }   } \,,
\een   
and return to the future singularity at  $r=0$.
Note that $r_{\rm max} \ge 2M$. 

For a recent treatment of photon orbits in the Schwarzschild metric
using Jacobi elliptic functions, the  reader is  directed to \cite{Munoz}.

\subsection{Properties of the Jacobi meric}
Since any spherically symmetric is conformally flat
we could adopt isotropic coordinates
and avail ourselves of the results given in \cite{Ong}.   
Alternatively, 
the calculation of the Gaussian curvature $K$ with respect
to the Jacobi metric  on the
equatorial planes  is straight
forward using equation (8) of \cite{Gibbons:2008hb}.  
However unless  $m=0$, this leads to
rather complicated and un-illuminating expresions. 
A qualititive analysis, based on the behaviour of
circular geodesics, i.e. those with $r=\const$   seems preferable.
We restrict attenton to an equatorial plane.

If $E^2  \ge m^2 $, $K$ tends zero at infinity
and in all three cases it tends to $ -\frac{1}{16M^2 E^2} $  
as $r$ tends to the horizon at  $r=2M$, which is at infinite Jacobi
radial distance,  the  curvature tends to a negative constant.
If $E^2 <m^2$, the Jacobi manifold has an  outer boundary 
at which the  metric  vanishes. This happens when  
\ben
E^2 = m^2 (1-\frac{2M}{r} ) \,, \qquad r= \frac{2Mm^2 } {m^2-E^2} > 2M  \,. 
\een 
This outer boundary should be thought of as a point since the Jacobi 
circumference
\ben
2 \pi \bigl (E^2 -m^2 (1-\frac{2M}{r}) \bigr ) \frac{r}
{\sqrt{1- \frac{2M}{r}}}\, 
\een
vanishes there.
In the vicinity of the boundary $K$ is positive.

Circular Jacobi  geodesics are possible. 
These correspond to extrema of the Jacobi circumference
and are located at values of $r$ for which
\ben
E^2 \bigl( 1-\frac{3M}{r} \bigr ) -m^2 \bigl(1-\frac{2M}{r}  \bigr )^2  =0 \,,\qquad \frac{E^2}{m^2}= \frac{(r-2M)^2}{r(r-2M)} \,.
\een
Circular geodsics exist with real energies
of all values of $r\ge 3M$. For every value of $\frac{E^2}{m^2}>1$ there is   
a unique circular geodesic with radius $r$  between $3M$ and $4M$. 
For every  value of $\frac{E^2}{m^2}$ between $\frac{8}{9} $ and $ 1$ 
there are two circular null geodesics, the inner,  which is unstable,
has its radius between $4M$ and $6M$, and the outer 
whose radius is greater than $6M$.   
Three interesting cases arise 
\begin{itemize}
\item  $ m^2 =0 \,, \qquad r=3M$. These are circular null geodsics
which are cirular geodsics of the optical metric.
\item $E^2 = m^2  \,, \quad r= 4M$, these are have the smalles raduius
among all circular   bound geodesics. 
\item $E^2= \frac{8}{9} m^2 \,,\qquad r=6M$. These are the most deeply bound
circular timelike geodesics.  
\end{itemize}

These results are consistent with
the   standard approach to circular timelike geodesics
which is  to  require    
the simultaneous vanishing of the right hand side  of (\ref{radial})
and its derivative. Eliminating $E^2$ and  solving for $u$ 
gives 
\ben
ru= \frac{1}{6M} \Bigl( 1  \pm \sqrt{1-\frac{12M^2}{h^2}} \Bigr ) \,.   
\een 
Elimininating $E$ and  solving  for $h$ gives
\ben
\frac{16 M^2 m^2}{h^2} = \frac{16(r-3M)}{r^2} \,.
\een  
For the unbound circular geodesics with $E^2>m^2$ , 
$\frac{16 M^2 m^2}{h^2}$ varies from
$0$ to $\frac{3}{4}$ as $r$ varies from $3M$ to $4M$.
For the bound circular geodsics with $ m^2 \le E^2 \le  \frac{8}{9} m^2 $
one finds that  
$\frac{16 M^2 m^2}{h^2}$ varies from  $\frac{3}{4}$ to $1$ as 
$r$ varies from $4M$ to $6M$ where it achieves its maximum value
and therafter as $r$ varies from $6M$ to infinity it decreases monotonically
to zero.

\subsection{Gauss Curvature and Isometric Embedding}
We can use these results to say something about the Gauss curvature $K$.

We begin by recalling that if the induced metric on the surface of revolution
$z=f(\rho)\,,\rho= \sqrt{x^2+y^2}$,is 
\ben
A^2 dr^2 + C^2(r) d \phi^2 = (1+ (f^\prime)^ 2 ) d \rho^2 + \rho ^2 d \phi ^2  
\een
then we have 
\ben
\rho= C(r) \,, \qquad (f^\prime) ^{-1}  = (\frac{1}{A} \frac{dC}{ dr})^2 -1) \,.\label{embedd}\een

In our case, the   embedding will extend from infinity towards the horizon
at $r=2M$ as long as the r.h.s of (\ref{embedd}) remains positive.
The Jacobi metric is conformal to
the spatial metric of the Schwarzschild solution and its
equatorial  plane is well known to be isometrically embeddable as
all the way down to the horizon as the Flamm paraboloid \cite{Flamm,Gibbons15}
\ben
z= \frac{( \rho -2M)^2}{8 M } \,.
\een  

However this is in general not possible for
the Jacobi metric. 

For example  if $m=0$  one is  limited us to  the region
$r> \frac{9}{4} M$. In that case
\ben
K= - \frac{2M}{r^3} (1- \frac{3M}{r}) 
\een 
which is everywhere negative and near the horizon the optical metric  
is asymptotic  to one of constant  negative curvature
equal to  $-\frac{1}{(4M)^2}$. 

This is analogous to the well known  fact that  the metric
with $A=1$ and $ C=ae^{-\frac{r}{a}}\,, a>0$ 
has constant  negative curvature
$-\frac{1}{a^2}$  and may be embedded into $\mathbb{E}^3$ 
as  the surface of revolution whose
meridional curve is a tractrix. However this  `` Beltrami's trumpet'' 
for which
\ben
f^\prime = \frac{1}{\rho} \sqrt{a^2-\rho^2}\,,\qquad \pm f= \const
+ a \cosh^{-1} (\frac{1}{\rho}) - \sqrt{a^2 - \rho ^2} \,,  
\een   
is incomplete and only occupies the region for which  $\rho <a$, i.e. $r>0$.

By (\ref{Sjacobi}) the  Jacobi metric is conformal to the optical metric
with conformal factor 
\ben
 E^2-m^2 + \frac{2M m^2 }{r} \,.
\een
It  also tends to constant negative curvature 
near the horizon.  Thus in general one does not expect to be able
to capture the near horizon geometry of the Jacobi metric by
an isometric embedding into $\mathbb{E}^3$ as a surface of revolution.
This represents   a practical  obstruction to constructing 
black hole analogues using such materials as graphene \cite{Cvetic:2012vg} .

Consider now the case  $m^2 > E^2 > \frac{8}{9} m^2 $.
Near the outer boundary the Gauss curvature is positive and
it is positive at the outer circular geodsic which is a  local
maximum of the Jacobi circumference. By the time we get to the
inner, unstable orbit, which is a  local
minimum  of the Jacobi circumference, the Gauss curvature is negative
and it is negative near the horizon at $r=2M$ near which the 
Jacobi circumference diverges. If
$  E^2 < \frac{8}{9} m^2 $ there are no circular geodesics
and the curvature is positive  near the boundary and 
negative near the horizon.  Thus the Gauss curvature of the Jacobi-metric
restricted to the equatorial plane is not everywhere negative as is
the case for the Fermat metric.

\section{The Jacobi metric for stationary spacetimes}

We cast the spacetime metric in Zermelo form \cite{Gibbons:2008zi}  
\ben
g_{\mu \nu}dx^\mu dx^\nu = \frac{V^2}{1-h_{ij}W^iW^j} \Bigl[-dt ^2 + h_{ij}(dx^i -W^i)(dx^j -W^j) \Bigr ]
\,,\label{smetric} \een
where I shall call  $h_{ij}$ the Zermelo metric, $W^i$ the wind,  and 
\ben
V^2 = -g_{\mu \nu} K^\mu K^\nu = - g_{tt} \,,
\een 
where $K^\mu\frac{\p}{\p x^\mu}= \frac{\p }{\p t}$,  is the timelike
Killing vector field. Note that if the wind vanishes, the
Zermelo metric coincides with the optical or Fermat metric. 

The   Lagrangian $L$ for a point particle of mass $m$ 
undergoing geodesic motion
in a  spacetime with metric (\ref{smetric}) is 
\ben
L= - \frac{mV}{\sqrt{1-h_{ij}W^iW^j}} 
 \sqrt{1- h_{ij}(\dot x^i -W^i)(\dot x^j - W^j) } 
\een  
where $\dot x^i= \frac{d x ^i}{dt}$.
The canonical momenta $p_i$ are therefore given by
\ben
p_i= \frac{mV}{\sqrt{1-h_{ij}W^iW^j} }  
\frac{h_{ij}( \dot x^j-W^j)  }{\sqrt{1- h_{ij}(\dot x^i -W^i)(\dot x^j - W^j)}     } \,. 
\een
The Hamiltonian, $H= p_i \dot x^i-L$, is given by
\ben
H(m,p_i,x^i) =     \sqrt{  h^{ij}p_ip_j  + \frac{m^2}{V^2}{1-h_{ij}W^iW^j} } + p_iW^i \,, \label{Ham}
\een 
where $h^{ik}h_{kj}= \delta ^i_j$.
Note that in the massles case, $m=0$  the Hamiltonian 
$H$ becomes
\ben
H (0,p_i,x^i)= \sqrt{  h^{ij}p_ip_j }    + p_iWi 
\een
which coincides with equation (14)  of \cite{Gibbons:2008zi}
thus recovering the result that the projection of
null geodesics of a stationary  spacetime  
onto the space of orbits of the timelike Killing vector   
solve the Zermelo problem of minimizing the time of travel 
in the presence of the wind and with respect to the 
Zermelo metric $h_{ij}$.

A quick way of obtaining (\ref{Ham}) 
is to work on the co-tangent bundle of spacetime.
The so-called super-Hamiltonian ${\cal H}$ is subject to the  constaint 
\ben
{\cal H} = g^{\mu \nu}p_\mu p_\nu = - m^2 \,. 
\label{super}
\een
If one solves (\ref{super}) for $p_0=-H$ 
one obtains (\ref{Ham}). 

In general, on the level set $H=E$  of the Hamiltonian we have 
\ben
h^{ij} p_ ip_j + \frac{m^2V^2}{(1-h_{ij}W^iW^j)} = (E-p_i W^i)^2.  
\een

If the mass $m$ is non-zero, it is not possible to cast this in he form
of an expression which is  a  homogeneous
degree two in momenta $p_i$ equated to a constant.
If it were so, then 
and  a Legendre transform would  
result in a Lagrangian which is of degree two in velocities
$v^=\dot x^i$ and hence we  would  be dealing with a 
Finsler structure, possibly Riemannian, as in the 
case of a static metric. Thus we are faced with a   geometric structure
more general than a Riemannian or even a  Finsler metric.

\end{document}